\documentclass [11pt, letterpaper]{article}
\usepackage{fullpage}
\usepackage{amsmath}
\usepackage{amssymb}
\usepackage{graphicx}
\usepackage{wrapfig}
\usepackage{boxedminipage}
\usepackage{epsfig}
\usepackage{setspace}
\usepackage{subfigure}
\usepackage{float}
\usepackage{hyperref}
\usepackage{soul}
\usepackage{enumerate}

\usepackage{cite}
\newcommand{\be}{\begin{equation}}
\newcommand{\bea}{\begin{eqnarray}}
\newcommand{\eea}{\end{eqnarray}}
\newcommand{\ba}{\begin{align}}
\newcommand{\ea}{\end{align}}
\newcommand{\ee}{\end{equation}}

\begin{document}

\begin{titlepage}
\vspace{10mm}
\begin{flushright}
 IPM/P-2013/044 \\
%FPAUO-12/10\\
\end{flushright}

\vspace*{20mm}
\begin{center}
{\Large {\bf Holographic  Entanglement Entropy for 4D Conformal Gravity }\\
}

\vspace*{15mm}
\vspace*{1mm}
{Mohsen Alishahiha${}^a$, Amin Faraji Astaneh$^{b,c}$ 
and  M. Reza Mohammadi Mozaffar$^a$ }

 \vspace*{1cm}

{\it ${}^a$ School of Physics, Institute for Research in Fundamental Sciences (IPM)\\
P.O. Box 19395-5531, Tehran, Iran \\  
${}^b$ Department of Physics, Sharif University of Technology,\\
P.O. Box 11365-9161, Tehran, Iran\\
${}^c $ School of Particles and Accelerators\\Institute for Research in Fundamental Sciences (IPM)\\
P.O. Box 19395-5531, Tehran, Iran \\ 
}
 \vspace*{0.5cm}
{E-mails: {\tt alishah@ipm.ir, faraji@ipm.ir, m$_{-}$mohammadi@ipm.ir}}%

\vspace*{2cm}
%%\maketitle
\end{center}

\begin{abstract}
Using the proposal for holographic entanglement entropy in higher 
derivative gravities, we compute holographic entanglement entropy for the conformal gravity in four dimensions which turns out to be finite. However, if one 
subtracts the contribution of the four dimensional Gauss-Bonnet term, the corresponding entanglement entropy
has a divergent term and indeed restricted to an Einstein solution of the conformal gravity,  the resultant 
entanglement entropy is exactly the same as that  in the Einstein gravity.
We will also make a comment on the first law of the  entanglement thermodynamics 
for the conformal gravity in four dimensions.
\end{abstract}

\end{titlepage}

%%%%%%%%%%%%%%%%%%%
%%%%%%%%%%%%%%%%%%%%%%%%%%%%%%%%%%%%%%%%%%%%%%%%%%%%%%%%%%%%%%%%%%%%%%%%%%%%%%
%%%%%%%%%%%%%%%%%%%%%%%%%%%%%%%%%%%%%%%%%%%%%%%%%%%%%%%%%%%%%%%%%%%%%%%%%%%%%%
\section{Introduction}
Four dimensional conformal gravity whose action is given in terms of the Weyl tensor, is a theory
that is invariant under a Weyl transformation of the metric. This, indeed, leads to a theory which depends
on angles, but not on distances.  Unlike the Einstein gravity\footnote{In this paper by ``Einstein gravity'' we mean an Einstein gravity with a cosmological constant.} which is ghost free, the conformal gravity has ghost, though it is 
renormalizable\cite{{Stelle:1976gc},{Adler:1982ri}}. It is also known that the solutions of  
Einstein gravity are solutions of
the conformal gravity too, though the inverse is not  necessarily  correct. It is, however, possible to 
impose a certain boundary condition on the metric 
in the conformal gravity, so that the theory admits only Einstein solutions\cite{Maldacena:2011mk}.

An interesting feature of the four dimensional conformal gravity is  its relation to the  
Einstein gravity in four dimensions. Actually, it was shown\cite{Anderson}
that the renormalized on shell action of a four dimensional Einstein gravity in an asymptotically hyperbolic Einstein spaces is given by the action of conformal gravity. Of course the action must be 
evaluated on an Einstein solution.  Moreover the author of 
\cite{Maldacena:2011mk} has  also argued  that the certain boundary condition, mentioned above, 
removes the ghost from the theory  and indeed the physical content of both theories would be the same.

Motivated by these observations, in the present paper, we will study the holographic entanglement 
entropy in the four dimensional conformal gravity with the aim of 
comparing the results with that of the  Einstein gravity. 

The action of the conformal gravity in four dimensions is 
\bea\label{action1}
S=-\frac{\kappa}{32\pi}\int d^{4}x\,\sqrt{-g}\,C_{\mu\nu\rho\sigma}C^{\mu\nu\rho\sigma}
=-\frac{\kappa}{32\pi}\int d^{4}x\,
\sqrt{-g}\,\bigg(R_{\mu\nu\rho\sigma}R^{\mu\nu\rho\sigma}-2 R_{\mu\nu}R^{\mu\nu}+\frac{1}{3}  R^2\bigg).
\eea
Here $\kappa$ is a dimensionless coupling constant which is the only free parameter of the  theory. The corresponding equations of motion, which is essentially the vanishing of the Bach tensor, are
\be
\left(\nabla^\mu\nabla^\nu-\frac{1}{2}R^{\mu\nu}\right)C_{\mu\rho\sigma\nu}=0.
\ee
These equations admit black hole solutions as follows\cite{Riegert:1984zz} (see also \cite{Grumiller:2010bz})
\be
ds^2=-F(r)\;dt^2+\frac{dr^2}{F(r)}+r^2d\Omega^2_2,\;\;\;\;\;\;\;
F(r)=\pm \sqrt{1-am}-\frac{m}{r}-\frac{ r^2}{L^2}+\frac{a r}{3},
\ee
where  $L$ is the radius of curvature and the $\pm$ signs refer to two branches of the solutions\cite{Maldacena:2011mk}. In what follows
we will only consider the plus sign, where the solutions are asymptotically AdS. It is also possible to consider the large volume limit\cite{Witten:1998zw}
so that the resultant solutions will be  black branes
\be\label{BB}
ds^2=\frac{L^2}{r^2}\bigg(-b(r)\;dt^2+\frac{dr^2}{b(r)}+\sum\limits_{i=1}^2 dx_i^2\bigg),
\;\;\;\;\;\;\;\;\;b(r)=1-\frac{a}{3}r\pm \sqrt{m a} r^2-mr^3.
\ee
Here $a$ and $m$ are the parameters of the solutions. Note that if one sets $a=0$ in the above solutions,
they reduce to AdS black hole (brane) solutions of the Einstein gravity in four dimensions.

On top of these solutions, we found another solution which is, indeed, an AdS wave solution (see
also \cite{{Alishahiha:2011yb},{Gullu:2011sj},{Lu:2012xu}}) 
\be\label{wave}
ds^2=\frac{L^2}{r^2}\bigg({dr^2}+dy^2-2dx_-dx_++k(x_+,r)\;dx_+^2\bigg),
\ee
with
\be
k(x_+,r)=c_0(x_+)+c_1(x_+)r+c_2(x_+)r^2+c_3(x_+)r^3.
\ee
For $c_1=c_2=0$ this solution reduces to an AdS wave solution in the Einstein gravity.
Note that since the Weyl transformation is a symmetry of the model, 
rescaling the above solutions with an arbitrary function leads to new metrics which are still solutions of the model. 

As we already mentioned the above solutions are not necessarily solutions of the Einstein 
gravity, though if one sets some of their  parameters to zero they reduce to solutions of 
the  Einstein gravity.  Therefore it is natural to pose a question whether there is a systematic way 
one could remove these terms from the solutions. Actually the answer is yes. To explore the situation note that  generally, using 
the Fefferman-Graham coordinates for a conformally asymptotically locally AdS solution,  the equations of motion
allow the following form for the metric
\be
ds^2=e^{\phi(r)}\frac{L^2}{r^2}\bigg(dr^2+g_{ij}(x,r)dx^idx^j\bigg),\;\;\;\;\;\;\;
g_{ij}(x,r)=g^{(0)}_{ij}+g^{(1)}_{ij}r+g^{(2)}_{ij}r^2+g^{(3)}_{ij}r^3+\cdots
\ee
It is then clear that by imposing a Neumann boundary condition as $\partial_r g_{ij}|=g_{ij}^{(1)}=0$ at the boundary
one can remove the linear term leading to an Einstein solution\cite{Maldacena:2011mk}. This is indeed the ghost
mode which is removed by the boundary condition.  It is clear that this condition sets $a=0$\footnote{
For the wave solution \eqref{wave} in order to get an Einstein solution one needs  to set both $c_1$ and $c_2$ to zero. But  imposing the Neumann boundary condition leads to $c_1=0$ and $c_2$ could still be non-zero.
More probably there should be another constrain on the solution to set either a relation between $c_1$ and $c_2$ or $c_2=0$, though we could not realize it. We should admit the resolution is not clear 
to us. 
Nevertheless since we are just interested in the Einstein solution ( and not the way we get it) one could 
simply set them to zero by hand.}.

It is worth noting that in the context of holographic  renormalization  it was argued that this extra
term would correspond to a source of a relevant operator in the dual conformal field theory\cite{{Grumiller:2013mxa},{Naseh}}.  More precisely, having non-zero linear term corresponds 
to deforming the corresponding dual theory with a relevant operator.

This is the aim of the present paper to compute the entanglement entropy of a three dimensional 
field  theory whose gravitational dual is given by the four dimensional conformal gravity \eqref{action1}. 
Since the action of the conformal gravity in four dimensions contains higher derivative terms, the simple
procedure of calculating the holographic entanglement entropy in terms of a minimal surface in the bulk\cite{Ryu:2006bv} is not applicable.
Therefore to compute the holographic entanglement entropy one should proceed with another method.

 Actually using a method based on the regularization of squashed cones 
the authors of \cite{Fursaev:2013fta} proposed an expression for the holographic 
entanglement entropy for actions with  curvature squared higher derivative terms. This proposal has been further studied in 
\cite{{Bhattacharyya:2013gra},{Alishahiha:2013zta}} for certain higher 
derivative gravities. A general formula for the holographic entanglement entropy for higher derivative gravities has been also proposed in 
\cite{{Dong:2013qoa},{Camps:2013zua}}. In what follows we will use the procedure of 
\cite{Fursaev:2013fta}  in which  the corresponding entropy functional for our case becomes
\begin{eqnarray}\label{EE}
S_A=\frac{ \kappa}{8} \int \;d^2\zeta \;\sqrt{h}\;\bigg[\frac{2}{3} R-2\left({ R}_{\mu\nu}n^\mu_i n^\nu_i-\frac{1}{2}\mathcal{K}^i\mathcal{K}_i\right)+2 \bigg( R_{\mu\nu\rho\sigma}n^\mu_i n^\nu_j n^\rho_i n^\sigma_j-\mathcal{K}^i_{\mu\nu}\mathcal{K}_i^{\mu\nu}\bigg)\bigg],
\end{eqnarray}
where $i=1,2$ denotes two transverse directions to a co-dimension two hypersurface in the bulk,
$n_i^\mu$  are two unit  mutually orthogonal normal vectors on the co-dimension two hypersurface 
and ${\cal K}^{(i)}$ is the trace of two extrinsic curvature tensors defined by
\be \label{ECUR}
{\cal K}_{\mu\nu}^{(i)}=\pi^\sigma_{\ \mu} \pi^\rho_{\ \nu}\nabla_\rho(n_i)_\sigma,\;\;\;\;\;\;\;
{\rm with}\;\;\;\;  \pi^\sigma_{\ \mu}=\epsilon^\sigma_{\ \mu}+\xi\sum_{i=1,2}(n_i)^\sigma(n_i)_\mu\ ,
\ee
where $\xi=-1$ for space-like and $\xi=1$ for time-like vectors. Moreover  $h$ is the 
induced metric on the hypersurface whose coordinates are denoted by $\zeta$.

Now the procedure is to consider an entangling region on the dual field theory on the boundary, then 
consider a co-dimension two hypersurface in the bulk whose intersection with the boundary coincides with the boundary of the entangling region. The profile of the co-dimension two hypersurface may be 
obtained by minimizing the above entropy functional. Moreover the entanglement entropy is the 
value of the entropy functional evaluated on the co-dimension two hypersurface.

The main results of our paper are as follows. We found that the entanglement entropy of theories whose
gravitational dual are provided by the four dimensional conformal gravity are finite and has no UV 
divergences which  usually appear as the effects of short term interactions. Nevertheless, if one subtracts the four dimensional Gauss-Bonnet action from the conformal action, as long as the
entanglement entropy is concerned,  the resultant action 
has the same physical content as that of the Einstein gravity.

The paper is organized as follows. In the next section we will study holographic entanglement entropy for 
black hole solutions in the conformal gravity in four dimensions where we observe that the resultant 
entanglement entropy is finite.  Moreover for the Einstein solution this finite part is exactly 
the same as that one finds in the  Einstein gravity using the 
minimal surface . In section three in order to explore our observation of the section two, we will 
study the entanglement entropy for the AdS plane wave solution of  the model. The last section is devoted to discussions.

%%%%%%%%%%%%%%%%%%%%%%%%%%%%%%%%%%%%%%%%%%%%%%%%%%%%%%%%%%%%%%%%%%
%%%%%%%%%%%%%%%%%%%%%%%%%%%%%%%%%%%%%%%%%%%%%%%%%%%%%%%%%%%%%%%%%%

\section{Entanglement entropy for black brane solutions}
%%%%%%%%%%%%%%%%%%%%%%%%%%%%%%%%%%%%%%%%%%%%%%%%%%%%%%%%%%%%%%%%%%
%%%%%%%%%%%%%%%%%%%%%%%%%%%%%%%%%%%%%%%%%%%%%%%%%%%%%%%%%%%%%%%%%%

In this section we study holographic entanglement entropy for a black brane solution in the 
four dimensional conformal gravity. We will consider the cases where the entangling region is 
either a strip or a disk.

\subsection{Strip entangling region}

In this subsection we calculate entanglement entropy for  an entangling region in the shape of a strip 
with the width of $\ell$. To do so, setting  $\sum\limits_{i=1}^2 dx_i^2=dx^2+dy^2$ in the equation
\eqref{BB} the entangling region may be given by 
\be\label{EnR}
-\frac{\ell}{2}\leq x\leq \frac{\ell}{2},\;\;\;\;\;\;\;\;\;0\leq y\leq L_y,\;\;\;\;\;\;\;\;t={\rm fixed}.
\ee
Then the corresponding  co-dimension two hypersurface in the bulk may be parametrized by  $t=0$ and $x=f(r)$
whose  induced metric becomes\footnote{Through out this paper we will be dealing with  three functions $f(r)$, $b(r)$ and 
$k(r)$ which are functions of $r$. But in order to simplify the expressions we will drop their
explicit dependence on $r$ and  will write them as $f$, $b$ and $k$, respectively.}

\begin{eqnarray}\label{inducedmetric}
ds^2=\frac{L^2}{r^2}\left[\left(\frac{1}{b}+f'^2\right)dr^2+dy^2\right].
\end{eqnarray}
Moreover the two unit vectors normal to the co-dimension two hypersurface are 
\begin{eqnarray}\label{sigma}
\Sigma_1 &:&\hspace*{0.5cm}t=0\hspace*{2.45cm}n_1=\frac{L\sqrt{b}}{r}(1,0,0,0)\nonumber\\
\Sigma_2 &:&\hspace*{0.5cm}x -f(r)=0\hspace*{1.2cm}n_2=\frac{L}{r \sqrt{1+bf'^2}}(0,-f',1,0).
\end{eqnarray}
Following the equation \eqref{ECUR} one can compute the extrinsic curvatures 
of the hypersurface. Indeed one gets
\be
{\cal K}_{\mu\nu}^{(1)}=0,\;\;\;\;\;\;\;\;\;\;
{\cal K}_{\mu\nu}^{(2)}=
  \left({\begin{array}{cccc}
   0 & 0 & 0&0 \\ 0 & b^{-1} A &  f' A & \\ 0 &  f'A &  b f'^2 A&0 \\
   0 & 0 &0 &B\\
  \end{array} } \right),
\ee
where 
\be
B=\frac{b L f'}{r^2 \sqrt{1+b f'^2}}, \;\;\;\;\;\;\;\;\;\;\;A=\frac{L\left[ \left(2 b-r b'\right) f'+2 b^2 f'^3-2 b r f''\right]}{2 r^2 \left(1+b f'^2\right)^{5/2}}.
\ee
%whose trace is
%\be
%{\cal K}^{(2)}=\frac{\left(4 b-r b'\right) f'+4 b^2 f'^3-2 b r f''}{2 L \left(1+b f'^2\right)^{3/2}}
%\ee
Using the expression for the extrinsic curvatures and the normal vectors one can compute the 
entropy functional \eqref{EE}. In fact in the present case one finds
\be\label{pp}
S_A=-\frac{ \kappa L_y}{4}\int dr\; \left[
\frac{\left(b' f'+2 b f''\right)^2}{4 \sqrt{b} \left(1+b f'^2\right)^{5/2}}+\frac{\left(2 b f'^2-1\right) b''}{3 \sqrt{b} \sqrt{1+b f'^2}}\right].
\ee
Now the aim is to minimize this entropy functional to find the profile of the hypersurface parametrized by
$f$ with conditions that $f$ is a smooth differentiable function and  $f(0)=\frac{\ell}{2}$. Before
going into details of minimizing procedure, we would like to make a comment on the form of 
the above entropy functional. 

Note that neither $L$ nor the radial coordinate $r$  appeared explicitly in the final form of the entropy
functional. This observation together with the fact that  both $b$ and $f$ are smooth differentiable functions,
leads to an interesting prediction on the form of the  entanglement entropy in this case.
Namely, since the integrand does not diverge at $r=0$, the resultant entanglement entropy does not have
 UV divergent terms. This is unlike the area formula in the Einstein gravity where the integrand has the following 
typical divergent form
\be
A\sim \int dr \frac{\sqrt{1+bf'^2}}{r^{d-1} \sqrt{b}} \;\;\;\;\;\;\;\stackrel{r\rightarrow 0}{\longrightarrow}\;\;\;\;\;\;\;
A\sim \int_\epsilon \frac{dr}{r^{d-1}},
\ee
where $\epsilon$ is a UV cut off. Therefore one may wonder how the conformal gravity would produce the Einstein gravity's results
once the Neumann boundary condition on the metric is imposed. This is, indeed, the aim of this 
subsection to address this question.

In order to minimize the above entropy functional one may proceed with the well known procedure
in the literature. Namely one may consider the entropy functional  as a one dimensional action 
whose Lagrangian is defined by $S_A=\int dr {\cal L}$. Therefore, in the present case, the corresponding 
equation of motion is
\be\label{eq1}
\frac{\partial^2}{\partial r^2}\left(\frac{\delta {\cal L}}{\delta f''}\right)-\frac{\partial}{\partial r}\left(\frac{\delta {\cal L}}{\delta f'}\right)+\frac{\delta {\cal L}}{\delta f}=0.
\ee
 Since the entropy functional \eqref{pp} is independent of $f$, one gets a conservation law which 
might be solved to find $f$. We note, however, that in general it is difficult to solve the resultant 
equation. Of course for pure AdS solution where $b=1$ there is an exact solution as follows
\be\label{AdS}
f'(r)=\frac{r^2}{\sqrt{r_t^4-r^4}}.
\ee
where $r_t$ is the turning point where $f'(r)\rightarrow\infty$. 

Then, one may expand the equation around the AdS solution for small deformations of parameters $m$ 
and $a$.  It is, however, important to note that since the equation of motion we get  for $f'(r)$ 
is a third order differential equation, in general it has three free parameters which should be
fixed by proper boundary conditions. The corresponding condition we will impose are finiteness and 
reality conditions on $\ell$. More precisely at leading order one finds
\be
f'(r)=\frac{r^2}{\sqrt{r_t^4-r^4}}\left[1+g(r)+\mathcal{O}(m^2,a^2,ma)\right],
\ee
with
\bea
g(r)&=&\frac{3 m r^5 \left(r^4-r_t^4\right)+ a r^3\left(r^4-2r_t^4\right)- \sqrt{a m} \left(3 r^8-4 r^4 r_t^4+r_t^8\right)}{6r^2\left(r^4-r_t^4\right)}+\frac{c_1}{r \left(r^4-r_t^4\right)}\nonumber\\
&+&\frac{c_2}{2 \left(r^4-r_t^4\right)}
+c_3\left(\frac{1}{12r^2 r_t^4\sqrt{r_t^4-r^4}}+\frac{E\left(\sin^{-1}\frac{r}{r_t},-1\right)- F\left(\sin^{-1}\frac{r}{r_t},-1\right)}{6r r_t^3\left(r_t^4-r^4\right)}\right).
\eea
where $E$ and $F$ are Elliptic functions. 
Using the fact that $\frac{\ell}{2}=\int_0^{r_t} dr f'(r)$ one can find a relation between  the width of the entangling region and the turning point. Of course the resultant width should be real and
finite. Indeed, requiring these  conditions, one   finds
\be
c_3=0,\;\;\;\;\;\;\;c_2=\frac{a}{3}r_t^5-\frac{2c_1}{r_t}.
\ee
Moreover  in the limit of $\ell\rightarrow 0$, the turning point must approach zero.
This condition would also require to set $c_1=0$. Therefore we arrive at\footnote{Note that 
in each order one needs to impose the reality and finiteness conditions on $\ell$.}
\begin{eqnarray*}\label{strip}
f'(r)&=&\frac{r^2}{\sqrt{r_t^4-r^4}}\left[1+\frac{1}{2}m r^3+\frac{a \left(r^5-2 r r_t^4+r_t^5\right)}{6\left(r^4-r_t^4\right)}+\frac{\sqrt{a m} (r_t^4-3r^4)}{6 r^2}+\mathcal{O}(m^2,a^2,ma)\right].
\end{eqnarray*}
As a result,  at leading order one gets
\begin{eqnarray}
\frac{\ell}{2}=\frac{\sqrt{\pi}\Gamma \left(\frac{3}{4}\right)}{\Gamma \left(\frac{1}{4}\right)}r_t
+\frac{m\pi }{16} r_t^4
+\frac{\sqrt{\pi}\Gamma \left(\frac{3}{4}\right)}{12\Gamma \left(\frac{1}{4}\right)} a r_t^2+\mathcal{O}(m^2,a^2,ma),
\end{eqnarray}
which can be inverted to find the turning point as a function of the width of the entangling region
\begin{eqnarray}\label{RT}
r_t=\frac{ \Gamma \left(\frac{1}{4}\right)}{2\sqrt{\pi } \Gamma \left(\frac{3}{4}\right)}\ell
-\frac{ \Gamma \left(\frac{1}{4}\right)^5}{256 \pi^{3/2}  \Gamma \left(\frac{3}{4}\right)^5}\;m \ell^4
-\frac{ \Gamma \left(\frac{1}{4}\right)^2}{48\pi  \Gamma \left(\frac{3}{4}\right)^2} \;a\ell^2
+\mathcal{O}(m^2,a^2,ma).
\end{eqnarray}
On the other hand plugging the profile \eqref{strip} into the entropy functional one arrives at
\begin{eqnarray}
S_{EE}=-\frac{\kappa L_y}{4}\left[\frac{4 \sqrt{\pi } \Gamma \left(\frac{3}{4}\right)}{ \Gamma \left(\frac{1}{4}\right)}\;\frac{1}{r_t}-\frac{m \pi}{2}   r_t^2
-\frac{ \sqrt{\pi } \Gamma \left(\frac{3}{4}\right)}{ 3\Gamma \left(\frac{1}{4}\right)}\;a 
\right]+\mathcal{O}(m^2,a^2,ma)
\end{eqnarray}
which, by making use of the equation \eqref{RT},  can be recast  to the following form 
\begin{eqnarray}\label{EEstrip}
S_{EE}= \kappa L_y\left[-\frac{2 \pi  \Gamma \left(\frac{3}{4}\right)^2}{ \Gamma \left(\frac{1}{4}\right)^2}\frac{1}{\ell}+\frac{\Gamma \left(\frac{1}{4}\right)^2}{64\Gamma \left(\frac{3}{4}\right)^2}\; m \ell^2\right]+\mathcal{O}(m^2,a^2,ma).
\end{eqnarray}
This is, indeed,  our final result for the holographic entanglement entropy for a black hole solution in the 
conformal gravity.  

An interesting feature of the  resultant entanglement entropy is that it  does not contain  UV divergent terms, as we had already anticipated. Moreover those finite terms which are independent of $a$, 
up to an overall factor, are exactly the same as that in the Einstein gravity\cite{Bhattacharya:2012mi}.
Therefore  setting $a=0$, the entanglement entropy reduces to that of Einstein gravity which can be obtained from the minimal area.

It is worth noting that although the entanglement entropy by definition is a positive
quantity, its finite term could be negative\cite{Bhattacharya:2012mi}. Therefore if with any procedure one 
removes the UV divergences of the theory, the resultant entanglement entropy could be negative.
Having found a finite negative entanglement entropy for the conformal gravity, it would mean that 
the corresponding  theory is intrinsically regularized. In what follows we will see this is, indeed, the case
and moreover we show how the divergences could be detached and removed.

In fact the finiteness of the entanglement entropy may be
understood from the fact that the  four dimensional conformal gravity could be considered as
a regularized four dimensional Einstein gravity \cite{Anderson}. So that all the divergences in the theory
should have  been removed. On the other hand following the results of \cite{Maldacena:2011mk} one would also
expect that setting $a=0$ where the solution becomes a solution of Einstein gravity, the
content of the model should also reduce to the Einstein gravity. 

To explore these points better, note that the action of the four dimensional conformal gravity may 
be decomposed as follows
\bea\label{decomp}
S&=&-\frac{\kappa}{32\pi}\int d^{4}x\,
\sqrt{-g}\,\bigg(R_{\mu\nu\rho\sigma}R^{\mu\nu\rho\sigma}-2 R_{\mu\nu}R^{\mu\nu}+\frac{1}{3} 
 R^2\bigg)\cr &&\cr
 &=&-\frac{\kappa}{32\pi}  {\rm GB}_4-\frac{\kappa}{16\pi} \int d^{4}x\,\sqrt{-g}\,\left(R_{\mu\nu}R^{\mu\nu}-\frac{1}{3}R^2\right),
\eea
where ${\rm GB}_4$ is the four dimensional Gauss-Bonnet action which is a total derivative and
does not contribute to the equations of motion. Note that since the Gauss-Bonnet term is topological, the
whole dynamics must be encoded in the second term. Therefore we will call the second term as the
``dynamical term''\footnote{Actually there are several ways to decompose the Weyl
action into a Gauss-Bonnet term plus a {\it dynamical term}. We note, however, that the
decomposition \eqref{decomp} is special in a sense that the coefficients of the Weyl
action and the Gauss-Bonnet action are the same. In other words the coefficient of the Gauss-Bonnet action
is one up to the factor of $\frac{\kappa}{32\pi}$. This is exactly the proper factor needed to regularize 
the four dimensional Einstein gravity with the Gauss-Bonnet term\cite{Miskovic:2009bm} (see also
\cite{{Maldacena:2011mk},{Anderson}}). Therefore the above dynamical terms is unique. As we will see, this 
leads to an interesting result concerning the connection between four dimensional conformal and Einstein gravities. }.  It is illustrative to compute the contributions
 of these two terms to the entanglement entropy, separately. 

To proceed let us consider the following action. 
\be
S^{{\rm dyn}}=-\frac{\kappa}{16\pi} \int d^{4}x\,\sqrt{-g}\,\left(R_{\mu\nu}R^{\mu\nu}-\frac{1}{3}R^2\right).
\ee
It is obvious that the black brane solution \eqref{BB} is also a solution of the equations of motion of the 
above action. Moreover, following \cite{Fursaev:2013fta} the holographic entanglement entropy of a dual field theory whose gravitational
description is given by the black brane solution of the above action must be obtained from the 
following entropy functional
\begin{eqnarray}\label{TT}
S^{{\rm dyn}}_A=\frac{\kappa}{4} \int \;d^2\zeta \;\sqrt{h}\;\bigg[-\frac{2}{3} R+\left({ R}_{\mu\nu}n^\mu_i n^\nu_i-\frac{1}{2}\mathcal{K}^i\mathcal{K}_i\right)\bigg].
\end{eqnarray}
For the entangling region \eqref{EnR} and its corresponding co-dimension two hypersurface  in the bulk, the
above entropy functional reads
\be\label{Dy}
S_A^{{\rm dyn}}=\frac{\kappa L_y}{8} \int \;dr\;\left[\frac{6 \left(3 b-r b'\right)-\left(1-2 b f'^2\right) \left(6 b-r^2 b''\right)}{3 \sqrt{b} r^2 \sqrt{1+b f'^2}}-\frac{\left(\left(4 b-r b'\right) f'+4 b^2 f'^3-2r b  f''\right)^2}{4 \sqrt{b} r^2 \left(1+b f'^2\right)^{5/2}} 
\right].
\ee
It is then straightforward to minimize the above entropy functional to read the corresponding 
profile of the hypersurface in the bulk. In fact solving the obtained equation perturbatively 
for small $a$ and $m$, we reach to the same profile as that in the equation \eqref{strip}. It is then 
easy to compute the entanglement entropy for the equation \eqref{TT} which  at   leading order is
\begin{eqnarray}
S^{{\rm dyn}}_{EE}=\kappa L_y\left[\frac{1}{\epsilon }-\frac{ a}{6}-\frac{2 \pi  \Gamma \left(\frac{3}{4}\right)^2}{\Gamma \left(\frac{1}{4}\right)^2}\;\frac{1}{\ell}+\frac{\Gamma \left(\frac{1}{4}\right)^2}{64\Gamma \left(\frac{3}{4}\right)^2}\;m\ell^2+...\right].
\end{eqnarray}
Going through the same procedure for the Gauss-Bonnet term, one arrives at
\begin{eqnarray}
S^{\rm GB}_{EE}= \kappa L_y\left[-\frac{1}{\epsilon}+\frac{a}{6}+...\right].
\end{eqnarray}
It is then clear that taking both contributions into account the divergent term will drop leading 
to a finite entanglement entropy in the equation \eqref{EEstrip}. Besides, setting $a=0$ where 
the solution becomes a  Schwarzschild  black brane,  the entanglement entropy
evaluated just by the dynamical part, $S_{EE}^{\rm dyn}$, reduces to the entanglement entropy of a
strip in the Einstein gravity if one identifies $\kappa$ as $\kappa=\frac{L^2}{2 G}$, 
 where $G$ is 
the four dimensional Newton constant\footnote{If we had 
considered another decomposition rather than \eqref{decomp} the result of the Gauss-Bonnet term would 
have been changed by an overall factor. Therefore the finite terms of the
entanglement entropy, for an Einstein solution, do not depend on the decomposition. Of course in this case the whole entanglement entropy obtained
just from the dynamical part would not be the same as that in the Einstein gravity.}. It is worth recalling  that setting $a=0$ corresponds to imposing the Neumann 
boundary condition on the metric which in turns reduces the solution to that of Einstein gravity\cite{Maldacena:2011mk}\footnote{To
 compare our normalization with that in \cite{Maldacena:2011mk}  one has $\kappa=32 \pi c_W$. See 
equation (3.6) of the paper \cite{Maldacena:2011mk}.}.

Actually we can show even more. Indeed the equation of motion obtained from the 
minimization of the equation \eqref{Dy} can be solved exactly leading to the following 
closed form for the profile of the hypersurface in the bulk
%\footnote{Note that this profile does not minimize the
%whole entropy functional where both dynamical part and Gauss-Bonnet term are taken into account.}
\be
f'(r)=\frac{r^2}{\sqrt{b\;(r_t^4-r^4)}},\;\;\;\;\;\;\;\;{\rm with}\;\;b=1-mr^3.
\ee
We recognize that this is exactly the same profile which minimizes the area, 
\be
S_A^{\rm Ein}=\frac{L^2 L_y}{4G}\int  dr \frac{\sqrt{1+b f'^2}}{r^2 \sqrt{b}},
\ee
 that is the entropy functional for the  Einstein gravity. Evaluating the entropy functional coming from  
the dynamical part and the area
function on this profile leads to the following  expressions for the holographic entanglement entropy
in the conformal and Einstein gravities, respectively
\be
S_{EE}^{\rm dyn}=\kappa L_y\int_\epsilon^{r_t}  dr\frac{ r_t^2}{r^2\sqrt{b \;(r_t^4-r^4)}},\;\;\;\;\;\;\;
S_{EE}^{\rm Ein}=\frac{L^2 L_y}{2G}\int_\epsilon^{r_t}  dr\frac{ r_t^2}{r^2\sqrt{b \;(r_t^4-r^4)}},
\ee
which are the same upon the identification of $\kappa=\frac{L^2}{2G}$.

Therefore it is fair to conclude that for an Einstein solution the holographic entanglement entropy 
evaluated by the dynamical part contains the same information as the holographic entanglement entropy
evaluated by the Einstein gravity where one has a simple proposal based on a minimal 
surface.  This conclusion may be understood as follows.  In fact 
by calculating the entanglement entropy, we are actually measuring the entanglement
between different degrees of freedom located on a given region. Therefore only dynamical modes
would contribute. On the other hand for the four dimensional conformal gravity the dynamics of the 
modes are governed by the dynamical part of the action. The topological Gauss-Bonnet term plays just
the role of a regulator which regularizes the results\footnote{
We note that the role of the Gauss-Bonnet term as a regulator
for AdS gravity action in four dimensions was also discussed in \cite{Olea:2005gb}.}.

In order to further explore this observation, in the following subsection we redo our calculations 
for the case where the entangling region is a disk.

%%%%%%%%%%%%%%%%%%%%%%%%%%%%%%%%%%%%%%%%%%%%%%%%%%%%%%%%%%%%%%%%%%

\subsection{Disk entangling region}

In this subsection we continue our studies on the holographic entanglement entropy of the
four dimensional conformal gravity for an entangling region in the shape of a disk. To proceed we reparametrize
the black brane metric as follows
\be
ds^2=\frac{L^2}{r^2}\left(-b\;dt^2+\frac{dr^2}{b}+d\rho^2+\rho^2 d\phi^2\right),\;\;\;\;\;\;\;\;\;
b=1-\frac{a}{3}r+\sqrt{am}r^2-m r^3.
\ee
Consider a disk on the boundary theory with the radius of $\ell$ given by  $\rho\leq \ell$, then the corresponding co-dimension two
hypersurface in the bulk may be parametrized by  $t=0,\rho=f(r)$.  Therefore the induced metric
on the hypersurface is 
\begin{eqnarray}\label{induced}
ds^2=\frac{L^2}{r^2}\left[\left(\frac{1}{b}+f'^2\right)dr^2+f^2 d\phi^2\right].
\end{eqnarray}
Moreover, two unit vectors normal to the hypersurface are also given by
\begin{eqnarray}\label{sigma1}
\Sigma_1 &:&\hspace*{0.5cm}t=0\hspace*{2.45cm}n_1=\frac{L\sqrt{b}}{r}(1,0,0,0)\nonumber\\
\Sigma_2 &:&\hspace*{0.5cm}\rho -f(r)=0\hspace*{1.2cm}n_2=\frac{L}{r \sqrt{1+bf'^2}}(0,-f',1,0).
\end{eqnarray}
It is then  easy to compute the extrinsic curvatures of the hypersurface associated to these vectors. 
Actually the results are the same as what we have found for the strip in the previous section, except that
in the present case one has
\be
B=\frac{f L \left(r+b f f'\right)}{r^2 \sqrt{1+b f'^2}}.
\ee
In this case  the  entropy functional, \eqref{EE}, reads
\be\label{Edisk}
S_A=-\frac{\pi \kappa}{4} \int dr\; \left[
\frac{\left(2+f b' f'+2 b f'^2+2 b f f''\right)^2}{4 \sqrt{b} f \left(1+b f'^2\right)^{5/2}}+
\frac{\left(2 b f'^2-1\right) f b''}{3 \sqrt{b} \sqrt{1+b f'^2}}
\right].
\ee
The entropy functional associated to the  dynamical part of the action can also be
computed leading to the following expression 
\be
S_A^{\rm dyn}=\frac{\pi \kappa}{4} \int dr \left[\frac{6 (3 b-r b')-(1-2 b f'^2)
 (6 b-r^2 b'')}{3 \sqrt{b}  f^{-1}r^2 \sqrt{1+b f'^2}}
-\frac{[2 (r+2 b f f') (1+b f'^2)-r f(b' f'+2 b f'')]^2}{4 \sqrt{b} f r^2 \left(1+b f'^2\right)^{5/2}}\right]
\ee
Now one needs to minimize the entropy functional to find a differential equation for the profile $f$.
Following our observation in the previous subsection, as long as  Einstein solutions are 
concerned one could only minimize the entropy functional associated to  the dynamical part
of action  to find the profile. Also the entropy can be obtained from this part.
 It is, however,  important to note that even for Einstein solutions the profile
we find by the  minimization of the entropy functional associated to the dynamical part
 does not necessarily minimize the whole entropy functional.  This means that the part of entropy functional which comes 
from the topological term, might have non-trivial effects on the solution of the profile. Nevertheless
as long as the finite parts of the entanglement entropy are  concerned both of them lead
to the same results for Einstein solutions. In what follows we will consider the entropy functional
of whole system where the effects of  both Gauss-Bonnet and dynamical parts are taken into account. 

Since in the  present case the entropy functional depends on $f$,  one does not have 
a conservation law and therefore the equation of motion has to be solved directly.  Of course in general it is difficult to solve  the resultant equation of motion. We note, however, that when the background is an AdS solution where $b=1$, the corresponding equation of motion admits an exact solution as follows
\be
f(r)=\sqrt{r_t^2-r^2},
\ee
which is exactly the same  as that in the Einstein gravity.  For the black brane solution, following our previous example, with proper boundary conditions  one can find a perturbative 
expansion for the profile for small $a$ and $m$ as follows
\begin{eqnarray}\label{profile1}
f(r)=\sqrt{r_t^2-r^2}&\bigg[&1+m\frac{2r_t^5-r^3\left(r_t^2+r^2\right)}{8\left(r_t^2-r^2\right)}-\frac{
\sqrt{ma} \left(\left(r+r_t\right) \left(r^2+2 r_t^2\right)- r_t^3 \tanh^{-1}\frac{r}{r_t}\right)}{6 \left(r+r_t\right)}+\nonumber\\
&&a\frac{r^2+r r_t+4 r_t^2}{12 \left(r+r_t\right)}\bigg]+\mathcal{O}(m^2,a^2, ma),
\end{eqnarray}
so that
\be
\ell=r_t\bigg[1+\frac{m}{4}r_t^3-\frac{
\sqrt{ma}}{3}  r_t^2+\frac{a}{3}r_t\bigg]+\mathcal{O}(m^2,a^2, ma),
\ee
which can be used to find the turning point as a function of the  radius of  the entangling region, $\ell$.
It easy to see that in the present case at leading order one just need to set  
$r_t=\ell$.

On the other hand form the entropy functional \eqref{Edisk} one gets
\begin{eqnarray}
S_{EE}=\frac{\pi \kappa}{8} mr_t^3+\mathcal{O}(m^2,a^2, ma),
\end{eqnarray}
so that 
\begin{eqnarray}
S_{EE}=\frac{\pi \kappa}{8} m\ell^3+\mathcal{O}(m^2,a^2, ma),
\end{eqnarray}
which is finite, as expected. It is worth nothing that  if one sets $m=0$ and $a=0$ where the 
solution reduces to pure AdS geometry, the finite part of the entanglement entropy vanishes.
It is unlike the entanglement entropy for a disk in Einstein gravity where the finite part is 
a universal constant\cite{Ryu:2006ef}. To explore this point better, it is illustrative to compute the
contributions of the dynamical part and the Gauss-Bonnet term to the entanglement entropy separately. Indeed for the 
dynamical part one finds
\be\label{dynSph}
S^{\rm dyn}_{EE}=\pi \kappa \left[\frac{\ell}{\epsilon }-1+\frac{m\ell^3}{8}-\frac{5 a \ell}{12}+\frac{\sqrt{ma} \ell^2 }{6}+\cdots\right],
\ee
while  the Gauss-Bonnet contribution to the entanglement entropy is
\be
S^{\rm GB}_{EE}=\pi \kappa \left[-\frac{\ell}{\epsilon }+1+\frac{5 a \ell}{12}-\frac{\sqrt{ma} \ell^2}{6} 
+\cdots\right].
\ee
As one observes both dynamical part and Gauss-Bonnet term contribute to the universal 
part but with opposite signs. Therefore the universal part drops when both contributions are
taken into account. 

It is also  clear that when one sets $a=0$ where the solution is a four dimensional Schwarzschild 
black brane of the Einstein gravity, the contribution of the  dynamical part is exactly the same
as the entanglement entropy obtained from minimal surface in the Einstein gravity if 
one identifies $\kappa=\frac{L^2}{2 G}$. The Gauss-Bonnet term plays the role of a regulator and its  contribution  removes the divergent term. 

%%%%%%%%%%%%%%%%%%%%%%%%%%%%%%%%%%%%%%%%%%%%%%%%%%%%%%%%%%%%%%%%%%
%%%%%%%%%%%%%%%%%%%%%%%%%%%%%%%%%%%%%%%%%%%%%%%%%%%%%%%%%%%%%%%%%%

\section{Entanglement entropy for wave solution}

%%%%%%%%%%%%%%%%%%%%%%%%%%%%%%%%%%%%%%%%%%%%%%%%%%%%%%%%%%%%%%%%%%
%%%%%%%%%%%%%%%%%%%%%%%%%%%%%%%%%%%%%%%%%%%%%%%%%%%%%%%%%%%%%%%%%%

In this section in order to further  explore the connection between holographic entanglement entropy in 
the four dimensional conformal gravity and that in the Einstein gravity we will study 
entanglement entropy for
a field theory whose gravitational description is given by the four dimensional conformal gravity 
on an AdS plane wave.  It is another  non-trivial example where the important
role of  the  dynamical part of the action may be also seen. 

As we have already mentioned, the equations of motion of the conformal gravity admit an AdS plane wave
solution as follows
\begin{eqnarray}
ds^2=\frac{L^2}{r^2}\bigg[dr^2-2dx_+dx_-+k(r)\;dx_+^2+dy^2\bigg],\;\;\;\;\;\;\;k(r)=c_0+c_1r+c_2r^2+c_3r^3.
\end{eqnarray}
The  constant $c_0$ can be set to zero by a shift. When $c_1$ and $c_2$ are non-zero this is
only a solution of conformal gravity. Since in what follows we are interested in 
the Einstein solution, we set $c_1=c_2=0$, so that  $k=mr^3$.
%This may be done by imposing the Neumann  boundary condition on the metric. 
 %Therefore in what follows we set

Let us  consider a strip in the dual theory whose width is extended along $y$ direction. More precisely 
one has
\be
(x_-,x_+)=(-z,z),\;\;\;\;\;\;\;\;\;\;\frac{\ell}{2}\leq y\leq \frac{\ell}{2},\;\;\;\;\;\;\;\;\;\;t={\rm fixed}.
\ee
Then the co-dimension two hypersurface in the bulk may be given by $y=f(r)$. Therefore two 
unit vectors normal to the hypersurface are
\begin{eqnarray}
&&\Sigma_1: x_++x_-=0 \hspace*{1cm}n_1=\frac{L}{r\sqrt{2+k}}(0,1,1,0)\nonumber\\
&&\Sigma_2: y-f(r)=0 \hspace*{1cm}n_2=\frac{L}{r\sqrt{1+f'^2}}(-f',0,0,1).
\end{eqnarray}
The  induced metric on the co-dimension two hypersurface in the bulk  becomes
\begin{eqnarray}
ds^2=\frac{L^2}{r^2}\bigg[(1+f'^2)dr^2+(2+k)dz^2\bigg].
\end{eqnarray}
Entanglement entropy for a field  theory whose dual is  the Einstein gravity on the above wave solution has been
studied  in \cite{Narayan:2012ks}. In this case, being Einstein gravity, one only needs to 
consider  the area of the hypersurface 
\be
S_{A}^{\rm Ein}=\frac{L^2 L_+}{4\sqrt{2} G}\int dr\frac{\sqrt{(2+k)(1+f'^2)}}{r^2}.
\ee
It is easy then to minimize this area to find the profile of the hypersurface in the bulk which is
\be\label{pro}
f'(r)=\frac{cr^2}{\sqrt{2+k-c^2 r^4}}.
\ee
where $c=\sqrt{2+k(r_t)}/r_t^2$, with $r_t$ being the turning point, is a constant of motion. Plugging the profile in the area, one can read the 
entanglement entropy 
\be\label{Einw}
S_{EE}^{\rm Ein}=\frac{L^2 L_+}{2\sqrt{2} G}\int_\epsilon^{r_t} dr\frac{ (2+k)}{r^2 \sqrt{2+k-c^2 r^4}}.
\ee

Now let us consider the case of conformal gravity. Following our results in the previous section 
for an Einstein solution the physical information  is  encoded in  entropy functional associated to the dynamical part of the action \eqref{TT}. For our entangling region the entropy 
functional associated to the  dynamical part reads
\be
S_A^{\rm dyn}=\frac{ \kappa L_+}{8\sqrt{2}}\int dr\bigg[
\frac{\sqrt{1+f'^2} (8+4 k-2 r k'+r^2 k'')}{r^2 \sqrt{2+k}}-\frac{[f' (1+f'^2) (-8-4 k+r k')+
2 r (2+k) f'']^2}{4 r^2 (2+k)^{3/2} (1+f'^2)^{5/2}}\bigg].
\ee
The profile of the hypersurface is obtained by minimizing this entropy functional. Although the
 equation of motion is lengthy, one can  easily verify that the profile
\eqref{pro} is still a solution of the corresponding equation. Therefore entanglement entropy can be 
calculated by plugging the profile \eqref{pro} into the above entropy functional. Doing so, one arrives at
\be
S_{EE}^{\rm dyn}=\frac{\kappa L_+}{\sqrt{2}}\int_\epsilon^{r_t}\frac{(2+k)}{r^2 \sqrt{2+k-c^2 r^4}},
\ee
which is exactly the same as that in the equation \eqref{Einw} with a proper identification of $\kappa$ as we have done 
in the previous section. Therefore, taking the results of \cite{Maldacena:2011mk} into account it is 
 natural to consider the Gauss-Bonnet term as  just a regulator which removes the divergency 
of the resultant entanglement entropy. Of course in order to explore the role of Gauss-Bonnet
term it is important to minimize the whole entropy functional when both the dynamical part and
Gauss-Bonnet term are taken into account.

For the entangling region we are considering  in this section, the entropy functional  \eqref{EE} reads
\begin{eqnarray}
S_{A}=-\frac{ \kappa L_+}{8\sqrt{2}}\int dr \bigg[\frac{k'^2 (4+f'^2)}
{4 (2+k)^{\frac{3}{2}} \sqrt{1+f'^2}}
+\frac{ (2+k) f''^2-k'f' (1+f'^2)f''+(1-f'^2) (1+f'^2)^2 k''}
{\sqrt{2+k} (1+f'^2)^{\frac{5}{2}}}\bigg].
\end{eqnarray}
It is then straightforward  to minimize this entropy functional to get a differential equation for the 
profile $f$.  It is worth noting that in the present case  the profile \eqref{pro} is not a solution of the 
obtained differential equation, even though we are considering an Einstein solution.

Therefore it is important to redo our computations for the whole entropy functional to 
explore the role of Gauss-Bonnet term. To proceed we will try to solve the differential  equation obtained
by minimizing the whole entropy functional perturbatively in power of $m$, for sufficiently small $m$.

When $m=0$, the solution reduces to a pure AdS solution. Thus the profile is just that of AdS case given 
in the equation \eqref{AdS}. On the other hand for small $m$, with proper boundary conditions ( as that 
in the previous examples), at leading order one finds
\begin{eqnarray}
f'(r)=\frac{r^2}{\sqrt{r_t^4-r^4}}\left[1+\frac{m r_t^4 (r^3-r_t^3)}{4   (r^4-r_t^4)}\right]+\cdots\ .
\end{eqnarray}
It is worth nothing that the above expression is, indeed, leading order contribution one may get 
from \eqref{pro} by expanding it for small $m$. Therefore the deviation from exact expression for the 
profile \eqref{pro} appears from higher orders. This expression may be integrated to get
\begin{eqnarray}
\frac{\ell}{2}=\frac{\sqrt{\pi }\Gamma \left(\frac{3}{4}\right)}{\Gamma \left(\frac{1}{4}\right)}r_t+\frac{m r_t^4}{16} \left(\pi -\frac{2\sqrt{\pi } \Gamma \left(\frac{3}{4}\right)}{\Gamma \left(\frac{1}{4}\right)}\right)+\cdots\, ,
\end{eqnarray}
which  can be inverted to find the turning point as a function of the width of strip as follows
\begin{eqnarray}
r_t=\frac{ \Gamma \left(\frac{1}{4}\right)}{2 \sqrt{\pi } \Gamma \left(\frac{3}{4}\right)}\ell+\frac{ \Gamma \left(\frac{1}{4}\right)^4 \left(-\sqrt{\pi } \Gamma \left(\frac{1}{4}\right)+2 \Gamma \left(\frac{3}{4}\right)\right)}{256 \pi ^2 \Gamma \left(\frac{3}{4}\right)^5}m\ell^4+\cdots\, .
\end{eqnarray}
It is then easy  to compute the contributions of the dynamical part and the Gauss-Bonnet term
to the entanglement entropy which are
\begin{eqnarray}
S^{GB}_{EE}&=&-\kappa L_+ \frac{1}{\epsilon}+\cdots\ ,\nonumber\\
S^{dyn.}_{EE}&=&\kappa L_+\left[\frac{1}{\epsilon}-\frac{2 \pi  \Gamma \left(\frac{3}{4}\right)^2}{ \Gamma \left(\frac{1}{4}\right)^2}\;\frac{1}{\ell}+\frac{  \Gamma \left(\frac{1}{4}\right)^2 }{64 \Gamma \left(\frac{3}{4}\right)^2}\;m\ell^2\right]+\cdots\ ,
\end{eqnarray}
that have the expected forms. Namely the Gauss-Bonnet term plays the role of a regulator and the
dynamical part reduces to that of Einstein gravity. 

%%%%%%%%%%%%%%%%%%%%%%%%%%%%%%%%%%%%%%%%%%%%%%%%%%%%%%%%%%%%%%%%%%
%%%%%%%%%%%%%%%%%%%%%%%%%%%%%%%%%%%%%%%%%%%%%%%%%%%%%%%%%%%%%%%%%%

\section{Discussions}
%%%%%%%%%%%%%%%%%%%%%%%%%%%%%%%%%%%%%%%%%%%%%%%%%%%%%%%%%%%%%%%%%%
%%%%%%%%%%%%%%%%%%%%%%%%%%%%%%%%%%%%%%%%%%%%%%%%%%%%%%%%%%%%%%%%%%
In this paper we have studied entanglement entropy of a quantum field theory whose
gravitational  description is provided by a four dimensional conformal gravity.  Since conformal gravities, typically, contain higher derivative terms, the 
simple holographic description of the entanglement entropy based on the 
minimal area is not applicable. 
Therefore in order to compute the entanglement  entropy we have used the prescription introduced in\cite{Fursaev:2013fta}. 
 
By making use of this method we have computed holographic entanglement entropy for a conformal gravity in four dimensions where we have  found that the resultant entanglement entropy, unlike 
the known examples in the literature,  is finite and has no UV divergences. 
The finiteness of the entanglement entropy may be understood from the fact that the 
Weyl action in four dimensions is equal to regularized 
on shell action of the Einstein gravity  when  all (classical) counter-terms in the bulk are  taken into account.
Therefore, using the holographic renormalization,  the UV divergences of the dual field theory which in turns correspond to the infinite volume limit in the bulk must be absent. Note that the finiteness of the entanglement entropy occurs both 
for Einstein and non-Einstein solutions. Of course, this is the case due to the fact that the finiteness is related to the regularization of the volume in the bulk and moreover,  since both solutions have the same 
asymptotic form, the volume regularization is the same for both of them.

Another interesting observation we have made is as follows.  Actually when one computes the holographic 
entanglement entropy for an Einstein solution of the conformal gravity the obtained 
entanglement entropy is exactly the same as the finite part of the entanglement entropy in the Einstein gravity. This might also be understood from the fact that the four dimensional  conformal gravity  
with certain  boundary condition, has the same physical content as that of the Einstein gravity
in  four dimensions. 

To explore this point better we note that  the action of the  four dimensional conformal gravity  can be decomposed into two parts.
The first part is just the four dimensional Gauss-Bonnet term which is topological and dose
not contribute to the equations of motion. The other part which we have called it ``dynamical part''
governs the dynamics of the system. As long as the equations of motion and their solutions are concerned the dynamical part plays the main role. Whereas, if we would like  to compute the thermal entropy or the energy of the solution the contribution of the Gauss-Bonnet term is essential as well.  Indeed in order to get the first law of the 
black hole thermodynamics it is important to consider the contribution of the Gauss-Bonnet term too.

On the other hand, in this paper by explicit examples, we have shown that for an Einstein solution the contribution 
of the dynamical part of the action to the entanglement entropy is exactly the same as that in the
Einstein gravity where the entanglement entropy is computed by minimizing the area of a co-dimension 
two hypersurface in the bulk. 
It is quite a non-trivial result, taking into account that in conformal gravity 
the entropy functional is not the area and indeed has rather a complicated expression. Note also that 
the obtained profile of the hypersurface in the bulk minimizes both the area and the entropy 
functional corresponding to the dynamical part of the conformal gravity. Although we have shown
this connection for certain entangling regions, it is natural to conjecture that this is, indeed, the case
for an arbitrary entangling region.
%\footnote{Note that this profile does not minimize the entropy functional of the conformal gravity 
%when both the Gauss-Bonnet and the dynamical part are taken into account.}.

Moreover,  as we have already  mentioned the entanglement entropy obtained from  total  action where 
the contribution of the Gauss-Bonnet is also taken into account is finite. This fact together with the 
above observation indicates that the Gauss-Bonnet term plays the role of a regulator which makes
the theory finite. It is then natural to imagine that the field theory dual to the four dimensional
conformal gravity, whatever it is, is finite. 

It is important to note that the above conclusion makes sense for those theories whose 
gravitational descriptions  are provided by conformal gravity on  asymptotically locally 
AdS solutions where the results of \cite{Anderson} is applied. Indeed, if one drops the assumption of 
being ``asymptotically hyperbolic'' for the metric the results of \cite{Anderson} fails  to 
hold and therefore there is no relation between Weyl action and regularized four dimensional
on shell Einstein action. Therefore the corresponding entanglement entropy is not finite. One can verify this
statement with an explicit example.

Actually the conformal gravity in four dimensions has the following  $z=4$
Lifshitz black hole 
solution\cite{Lu:2012xu}
\be
ds^2=\frac{L^2}{r^2}\left[-\frac{b}{r^6}dt^2+\frac{dr^2}{r^2b}+\sum_{i=1}^2dx_i^2\right],\;\;\;\;\;\;\;\;\;
b=1+c_1 r^2+\frac{c_1^2}{3}r^4+c_2 r^6,
\ee
which is not asymptotically AdS solution. It is then easy to compute the holographic entanglement entropy for this solution. Indeed  setting $b=1$  (for simplicity) the entropy functional \eqref{EE}  for the strip \eqref{EnR}
reads
\be
S_A=\frac{\kappa L_y}{8}\int dr\;\left[\frac{8-8 f'^4 (3+2 f'^2)-r^2 f''^2}{r^2 (1+f'^2)^{5/2}}\right],
\ee
which leads to the  following UV divergent term in the entanglement entropy
\be
S_{EE}=\kappa L_y\;\frac{1}{\epsilon}+\cdots\ .
\ee
One observes that,  even though,  the Gauss-Bonnet term is also taken into account the result is not 
UV finite. More precisely doing the same for  dynamical and Gauss-Bonnet parts one finds
\be
S_{EE}^{{\rm dyn}}=\frac{3\kappa L_y}{2}\;\frac{1}{\epsilon}+\cdots\;\;\;\;\;\;\;\;\;\;\;\;\;\;
 S_{EE}^{\rm GB}=-\frac{\kappa L_y}{2}\;\frac{1}{\epsilon}+\cdots\ .
\ee

As a side comment, note that from our computations one may calculate the variation of entanglement
entropy when  the system changes from a ground state to an excited state. Besides, since 
in all cases we have considered that the metric is asymptotically AdS, it is possible to compute the
expectation value of the energy momentum tensor of the dual field theory using holographic 
renormalization. Therefore it is possible to verify whether the resultant variation of energy and 
entanglement entropy satisfy the first law of the entanglement thermodynamics\cite{{Bhattacharya:2012mi},{Allahbakhshi:2013rda},{Blanco:2013joa},{Wong:2013gua}}.
Actually one finds that for an Einstein solution, the variation 
of the entanglement entropy and energy satisfy the first law of the entanglement thermodynamics with  the same entanglement temperature
as that in the Einstein gravity. Indeed, unlike the first law of black hole thermodynamics, the Gauss-bonnet term has no contribution. 

We note, however, that for  non-Einstein solutions (for example when $a\neq 0$) the
Gauss-Bonnet term does contribute. For example if one looks at the expression of $S_{EE}^{\rm dyn}$
for the sphere, \eqref{dynSph} and try to read the first law from this paper, due to $\sqrt{ma}$-term one finds an extra term in the first law of the
entanglement thermodynamics proportional to $\sqrt{a}$.
On the other hand  if one  reads the first law from the total
entanglement entropy due to the contribution of the Gauss-Bonnet the extra term drops from the
equation, leading to the same relation as that of the Einstein gravity. 

\
%%%%%%%%%%%%%%%%%%%%%%%%%%%%%%%%%%%%%%%%%%%%%%%%%%%%%%
%%%%%%%%%%%%%%%%%%%%%%%%%%%%%%%%%%%%%%%%%%%%%%%%%%%%%%
  
\section*{Acknowledgements}

We would like to thank  A. Naseh for a discussion.
We also acknowledge the use of M. Headrick's excellent Mathematica package "diffgeo". We would like
to thank him for his generosity. We would also like to thank the referee for his/her useful comments.

%%%%%%%%%%%%%%%%%%%%%%%%%%%%%%%%%%%%%%%%%%%%%%%%%%%%%%
%%%%%%%%%%%%%%%%%%%%%%%%%%%%%%%%%%%%%%%%%%%%%%%%%%%%%

\


\begin{thebibliography}{}




%\cite{Stelle:1976gc}{Adler:1982ri}
\bibitem{Stelle:1976gc} 
  K.~S.~Stelle,
  ``Renormalization of Higher Derivative Quantum Gravity,''
  Phys.\ Rev.\ D {\bf 16}, 953 (1977).
  %%CITATION = PHRVA,D16,953;%%
  %922 citations counted in INSPIRE as of 08 Nov 2013



%\cite{Adler:1982ri}
\bibitem{Adler:1982ri} 
  S.~L.~Adler,
  ``Einstein Gravity as a Symmetry Breaking Effect in Quantum Field Theory,''
  Rev.\ Mod.\ Phys.\  {\bf 54}, 729 (1982)
  [Erratum-ibid.\  {\bf 55}, 837 (1983)].
  %%CITATION = RMPHA,54,729;%%
  %465 citations counted in INSPIRE as of 08 Nov 2013


%\cite{Maldacena:2011mk}{Anderson}
\bibitem{Maldacena:2011mk} 
  J.~Maldacena,
  ``Einstein Gravity from Conformal Gravity,''
  arXiv:1105.5632 [hep-th].
  %%CITATION = ARXIV:1105.5632;%%
  %80 citations counted in INSPIRE as of 08 Nov 2013



\bibitem{Anderson}

M.~T.~ Anderson, ``$L^2$ curvature and volume renormalization of AHE metrics on 4-manifolds,''
Math.\ Res.\ Lett.,\ {\bf 8}  171 (2001) [ arXiv:math/001105].
 %%CITATION =math/001105;%%



%\cite{Riegert:1984zz}
\bibitem{Riegert:1984zz} 
  R.~J.~Riegert,
  ``Birkhoff's Theorem in Conformal Gravity,''
  Phys.\ Rev.\ Lett.\  {\bf 53}, 315 (1984).
  %%CITATION = PRLTA,53,315;%%
  %52 citations counted in INSPIRE as of 14 Nov 2013


%\cite{Grumiller:2010bz}
\bibitem{Grumiller:2010bz} 
  D.~Grumiller,
  ``Model for gravity at large distances,''
  Phys.\ Rev.\ Lett.\  {\bf 105}, 211303 (2010)
  [Erratum-ibid.\  {\bf 106}, 039901 (2011)]
  [arXiv:1011.3625 [astro-ph.CO]].
  %%CITATION = ARXIV:1011.3625;%%
  %31 citations counted in INSPIRE as of 14 Nov 2013








%\cite{Witten:1998zw}
\bibitem{Witten:1998zw} 
  E.~Witten,
  ``Anti-de Sitter space, thermal phase transition, and confinement in gauge theories,''
  Adv.\ Theor.\ Math.\ Phys.\  {\bf 2}, 505 (1998)
  [hep-th/9803131].
  %%CITATION = HEP-TH/9803131;%%

%\cite{Alishahiha:2011yb}{Gullu:2011sj}
\bibitem{Alishahiha:2011yb} 
  M.~Alishahiha and R.~Fareghbal,
  ``D-Dimensional Log Gravity,''
  Phys.\ Rev.\ D {\bf 83}, 084052 (2011)
  [arXiv:1101.5891 [hep-th]].
  %%CITATION = ARXIV:1101.5891;%%
  %41 citations counted in INSPIRE as of 14 Nov 2013


%\cite{Gullu:2011sj}
\bibitem{Gullu:2011sj} 
  I.~Gullu, M.~Gurses, T.~C.~Sisman and B.~Tekin,
  ``AdS Waves as Exact Solutions to Quadratic Gravity,''
  Phys.\ Rev.\ D {\bf 83}, 084015 (2011)
  [arXiv:1102.1921 [hep-th]].
  %%CITATION = ARXIV:1102.1921;%%
  %27 citations counted in INSPIRE as of 14 Nov 2013


%\cite{Lu:2012xu}
\bibitem{Lu:2012xu} 
  H.~Lu, Y.~Pang, C.~N.~Pope and J.~F.~Vazquez-Poritz,
  ``AdS and Lifshitz Black Holes in Conformal and Einstein-Weyl Gravities,''
  Phys.\ Rev.\ D {\bf 86}, 044011 (2012)
  [arXiv:1204.1062 [hep-th]].
  %%CITATION = ARXIV:1204.1062;%%
  %31 citations counted in INSPIRE as of 14 Nov 2013









%\cite{Grumiller:2013mxa}{Naseh}
\bibitem{Grumiller:2013mxa} 
  D.~Grumiller, M.~Irakleidou, I.~Lovrekovic and R.~McNees,
  ``Conformal gravity holography in four dimensions,''
  arXiv:1310.0819 [hep-th].
  %%CITATION = ARXIV:1310.0819;%%

\bibitem{Naseh}
A. Naseh, {\it unpublished}.


%\cite{Ryu:2006bv}
\bibitem{Ryu:2006bv} 

  S.~Ryu and T.~Takayanagi,
  ``Holographic derivation of entanglement entropy from AdS/CFT,''
  Phys.\ Rev.\ Lett.\  {\bf 96}, 181602 (2006)
  [hep-th/0603001].
  %%CITATION = HEP-TH/0603001;%%




\bibitem{Fursaev:2013fta} 
  D.~V.~Fursaev, A.~Patrushev and S.~N.~Solodukhin,
  ``Distributional Geometry of Squashed Cones,''
  arXiv:1306.4000 [hep-th].
  %%CITATION = ARXIV:1306.4000;%%




  %\cite{Bhattacharyya:2013gra}
\bibitem{Bhattacharyya:2013gra} 
  A.~Bhattacharyya, M.~Sharma and A.~Sinha,
  ``On generalized gravitational entropy, squashed cones and holography,''
  arXiv:1308.5748 [hep-th].
  %%CITATION = ARXIV:1308.5748;%%
  %1 citations counted in INSPIRE as of 18 Sep 2013
  

%\cite{Alishahiha:2013zta}
\bibitem{Alishahiha:2013zta} 
  M.~Alishahiha, A.~F.~Astaneh and M.~R.~M.~Mozaffar,
  ``Entanglement Entropy for Logarithmic Conformal Field Theory,''
  arXiv:1310.4294 [hep-th].
  %%CITATION = ARXIV:1310.4294;%%
  %1 citations counted in INSPIRE as of 08 Nov 2013


%\cite{Dong:2013qoa}{Camps:2013zua}
\bibitem{Dong:2013qoa} 
  X.~Dong,
  ``Holographic Entanglement Entropy for General Higher Derivative Gravity,''
  arXiv:1310.5713 [hep-th].
  %%CITATION = ARXIV:1310.5713;%%
  %2 citations counted in INSPIRE as of 08 Nov 2013


%\cite{Camps:2013zua}
\bibitem{Camps:2013zua} 
  J.~Camps,
  ``Generalized entropy and higher derivative Gravity,''
  arXiv:1310.6659 [hep-th].
  %%CITATION = ARXIV:1310.6659;%%


%\cite{Bhattacharya:2012mi}
\bibitem{Bhattacharya:2012mi} 
  J.~Bhattacharya, M.~Nozaki, T.~Takayanagi and T.~Ugajin,
  ``Thermodynamical Property of Entanglement Entropy for Excited States,''
  Phys.\ Rev.\ Lett.\  {\bf 110}, no. 9, 091602 (2013)
  [arXiv:1212.1164].
  %%CITATION = ARXIV:1212.1164;%%
  %20 citations counted in INSPIRE as of 09 Nov 2013

%\cite{Miskovic:2009bm}
\bibitem{Miskovic:2009bm} 
  O.~Miskovic and R.~Olea,
  ``Topological regularization and self-duality in four-dimensional anti-de Sitter gravity,''
  Phys.\ Rev.\ D {\bf 79}, 124020 (2009)
  [arXiv:0902.2082 [hep-th]].
  %%CITATION = ARXIV:0902.2082;%%
  %22 citations counted in INSPIRE as of 22 Dec 2013

%\cite{Olea:2005gb}
\bibitem{Olea:2005gb} 
  R.~Olea,
  ``Mass, angular momentum and thermodynamics in four-dimensional Kerr-AdS black holes,''
  JHEP {\bf 0506}, 023 (2005)
  [hep-th/0504233].
  %%CITATION = HEP-TH/0504233;%%
  %59 citations counted in INSPIRE as of 22 Dec 2013

%\cite{Ryu:2006ef}
\bibitem{Ryu:2006ef} 
  S.~Ryu and T.~Takayanagi,
  ``Aspects of Holographic Entanglement Entropy,''
  JHEP {\bf 0608}, 045 (2006)
  [hep-th/0605073].
  %%CITATION = HEP-TH/0605073;%%
  %282 citations counted in INSPIRE as of 10 Nov 2013

%\cite{Narayan:2012ks}
\bibitem{Narayan:2012ks} 
  K.~Narayan, T.~Takayanagi and S.~P.~Trivedi,
  ``AdS plane waves and entanglement entropy,''
  JHEP {\bf 1304}, 051 (2013)
  [arXiv:1212.4328 [hep-th]].
  %%CITATION = ARXIV:1212.4328;%%
  %7 citations counted in INSPIRE as of 21 Oct 2013


%\cite{Allahbakhshi:2013rda}{Wong:2013gua}
\bibitem{Allahbakhshi:2013rda} 
  D.~Allahbakhshi, M.~Alishahiha and A.~Naseh,
  ``Entanglement Thermodynamics,''
  JHEP {\bf 1308}, 102 (2013)
  [arXiv:1305.2728 [hep-th]].
  %%CITATION = ARXIV:1305.2728;%%
  %13 citations counted in INSPIRE as of 12 Nov 2013

%\cite{Blanco:2013joa}{Allahbakhshi:2013rda}{Wong:2013gua}
\bibitem{Blanco:2013joa} 
  D.~D.~Blanco, H.~Casini, L.~-Y.~Hung and R.~C.~Myers,
  ``Relative Entropy and Holography,''
  JHEP {\bf 1308}, 060 (2013)
  [arXiv:1305.3182 [hep-th]].
  %%CITATION = ARXIV:1305.3182;%%
  %14 citations counted in INSPIRE as of 12 Nov 2013

%\cite{Wong:2013gua}
\bibitem{Wong:2013gua} 
  G.~Wong, I.~Klich, L.~A.~Pando Zayas and D.~Vaman,
  ``Entanglement Temperature and Entanglement Entropy of Excited States,''
  arXiv:1305.3291 [hep-th].
  %%CITATION = ARXIV:1305.3291;%%
  %11 citations counted in INSPIRE as of 12 Nov 2013








\end{thebibliography}
\end{document}